\pdfoutput=1
\documentclass[aps,prl,twocolumn,groupedaddress,floatfix]{revtex4}

\usepackage{graphicx} 
\usepackage{mathptmx} 
\usepackage{amssymb} 
\usepackage{textcomp} 
\usepackage{helvet} 

\newcommand{\ad}{a^{\dag}}
\newcommand{\bd}{b^{\dag}}

\newcommand{\ketdown}{|\downarrow\rangle}
\newcommand{\ketup}{|\uparrow\rangle}
\newcommand{\on}{\omega_0}
\newcommand{\Oex}{\Omega_\mathrm{ex}}

\newcommand{\tex}{\tau_\mathrm{ex}}

\begin{document}

\title{Coupled quantized mechanical oscillators}

\author{K. R. Brown}
\email{kenton.brown@nist.gov}
\author{C. Ospelkaus}
\author{Y. Colombe}
\author{A. C. Wilson}
\author{D. Leibfried}
\author{D. J. Wineland}
\affiliation{Time and Frequency Division, National Institute of Standards and Technology, 325 Broadway, Boulder, CO 80305, USA}

\maketitle

\noindent\textbf{
  The harmonic oscillator is one of the simplest physical systems but
  also one of the most fundamental. It is ubiquitous in nature, often
  serving as an approximation for a more complicated system or as a
  building block in larger models. Realizations of harmonic
  oscillators in the quantum regime include electromagnetic fields in
  a cavity~\cite{haroche06,miller05,houck07} and the mechanical modes
  of a trapped atom~\cite{leibfried03} or macroscopic
  solid~\cite{oconnell10}.  Quantized interaction between two motional
  modes of an individual trapped ion has been achieved by coupling
  through optical fields~\cite{monroe01}, and entangled motion of two
  ions in separate locations has been accomplished indirectly through
  their internal states~\cite{jost09}. However, direct controllable
  coupling between quantized mechanical oscillators held in separate
  locations has not been realized previously. Here we implement such
  coupling through the mutual Coulomb interaction of two ions held in
  trapping potentials separated by 40~$\mu$m (similar work is reported
  in a related paper~\cite{harlander11}). By tuning the confining wells into
  resonance, energy is exchanged between the ions at the quantum
  level, establishing that direct coherent motional coupling is
  possible for separately trapped ions. The system demonstrates a
  building block for quantum information processing and quantum
  simulation. More broadly, this work is a natural precursor to
  experiments in hybrid quantum systems, such as coupling a trapped
  ion to a quantized macroscopic mechanical or electrical
  oscillator~~\cite{heinzen90,wineland98,tian04,hensinger05,tian05}.  }

The direct coupling of atomic ions in separate potential wells is a
key feature of proposals to implement quantum
simulation~\cite{schmied08,chiaverini08, schmied09}, and it could
allow logic operations to be performed in a multi-zone quantum
information processor~\cite{wineland98,cirac00,kielpinski02} without
the requirement of bringing the ion qubits into the same trapping
potential. Moreover, the coupling could prove useful for metrology and
sensing. For example, it could extend the capabilities of
quantum-logic spectroscopy~\cite{heinzen90,schmidt05,rosenband08} to
ions that cannot be trapped within the same potential well as the
measurement ion, such as oppositely charged ions or even anti-matter
particles~\cite{heinzen90,wineland98}.  Coupling could be obtained
either through mutually shared electrodes~\cite{heinzen90,daniilidis09}
or directly through the Coulomb
interaction~\cite{wineland98,cirac00,tan02,ciaramicoli03}.

The Coulomb interaction potential for
two trapped ions, $a$ and $b$, with charges $q_a$ and $q_b$ in potential
wells separated by a distance $s_0$ is given by
\begin{eqnarray*}
U(x_a,x_b)&=&\frac{1}{4 \pi \epsilon_0}\frac{q_aq_b}{s_0-x_a+x_b}\\
&\approx&\frac{1}{4 \pi
  \epsilon_0}\frac{q_aq_b}{s_0}(1+\frac{x_a-x_b}{s_0}+\frac{x_a^2}{s_0^2}
+ \frac{x_b^2}{s_0^2} - \frac{2x_ax_b}{s_0^2}) .\nonumber
\end{eqnarray*}
Here $x_a$ and $x_b$ are the displacements of the ions from the
external potential minima and $\epsilon_0$ is the permittivity of free
space. The first term is constant and does not affect the
dynamics. The second term represents a steady force between the ions
that displaces them slightly; if necessary, it can be counteracted
with additional potentials applied to nearby electrodes. The terms
proportional to $x_a^2$ and $x_b^2$ represent static changes in
the trap frequencies that could also be compensated with potentials
applied to nearby electrodes. The term proportional to $x_ax_b$ represents the
lowest-order coupling between the ions' motions. For small deviations,
$x'_a$ and $x'_b$, from equilibrium, the coupling is
\begin{equation}
\frac{-q_aq_b}{2 \pi \epsilon_0 s_0^3} (x'_ax'_b) = -\hbar \Oex (a+a^{\dag})(b+b^{\dag}) \approx -\hbar \Oex  (ab^{\dag} + a^{\dag}b), \label{eqn:interaction}
\end{equation}
where
\begin{equation}
\Oex \equiv \frac{q_aq_b}{4 \pi \epsilon_0 s_0^3 \sqrt{m_am_b}\sqrt{\omega_{0a}\omega_{0b}}}, \label{eqn:interaction_strength}
\end{equation}
and $a$, $a^{\dag}$, $b$, and $b^{\dag}$ represent the harmonic
oscillator lowering ($a$, $b$) and raising ($a^{\dag}$, $b^{\dag}$)
operators, $m_i$ and $\omega_{0i}$ are respectively the ion masses and
motional frequencies, $\hbar$ is Planck's constant divided by $2\pi$,
and we have neglected fast-rotating terms. Minimizing the distance,
$s_0$, between ions is crucial, because for fixed $\omega_{0i}$, the
coupling rate scales as $\Oex
\propto 1/s_0^3$.

When $\omega_{0a} = \omega_{0b} = \on$ (the resonance condition), we find
\begin{eqnarray}
\ad(t)&=&\exp(i \on t) (\ad(0)\cos(\Oex t) - i\bd(0)\sin(\Oex t))
\label{eqn:sol}\\
\bd(t)&=&\exp(i \on t) (\bd(0)\cos(\Oex t) - i\ad(0)\sin(\Oex t)). \nonumber
\end{eqnarray}
At time $t = \tex \equiv \pi/2\Omega_\mathrm{ex}$, the
operators have changed roles up to a phase factor, and the oscillators
have completely swapped their energies, regardless of their initial
states. At $t=2\tex$, the energies have returned to their initial
values in each ion. The mean occupation, $\langle \ad a \rangle$,
of ion $a$ as a function of time exhibits
oscillations with period $2 \tex$.

\begin{figure}
\centering \includegraphics{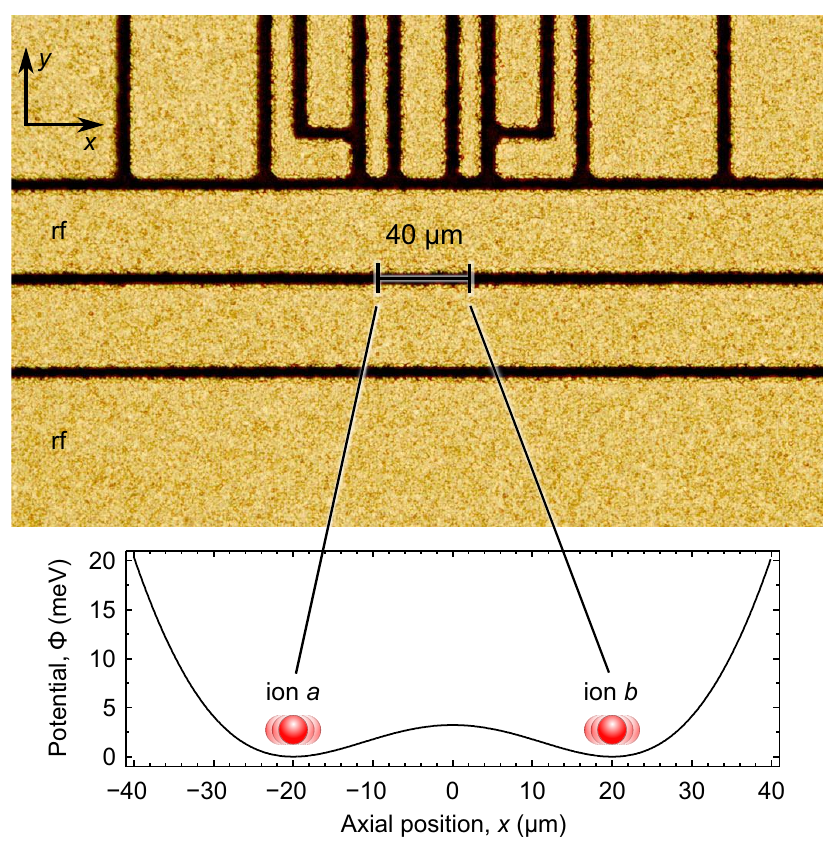}
 \sffamily \caption{Micrograph of the ion trap, showing
 radio-frequency (RF) and d.c.  electrodes, and gaps between
 electrodes (darker areas). The lower part of the figure indicates the
 simulated potential along the trap $x$ axis. Two trapping wells are
 separated by 40~$\mu$m, with ion positions marked by red spheres. The
 d.c. electrodes are sufficient to control the axial frequency and the
 position of each ion independently. Here both frequencies are
 $\sim4$~MHz and the potential barrier between the two ions is
 $\sim3$~meV.}
\label{fig:trap_center}
\end{figure}

Figure~\ref{fig:trap_center} shows a micrograph of our
surface-electrode trap~\cite{seidelin06}, constructed of gold
electrodes, 8~$\mu$m thick with 5-$\mu$m gaps, electroplated onto a
crystalline quartz substrate.
The trap can produce two potential minima at a height $d_0=40$~$\mu$m
above the surface and separated by $s_0=40\ \mu$m along the $x$
axis. Each potential well confines a single $^9\mathrm{Be}^+$ ion with
an axial (parallel to $x$) motional frequency of
$\omega_0/(2\pi)\approx4$~MHz and a barrier between wells of
$\sim3$~meV. Pseudopotential confinement in radial directions (normal
to $x$) is accomplished with a peak potential of $\sim100$~V at
170~MHz applied to the radio-frequency electrodes, yielding radial
frequencies of $\sim22$~MHz. By applying static potentials to the
d.c. electrodes, we can independently vary the separation between the
ions and the curvatures of the two trapping wells. In this way, the
ion axial motional frequencies can be brought into or out of
resonance, allowing a tunable interaction. For
$\omega_0/2\pi=4.04$~MHz, we predict that $\tex=162$~$\mu\mathrm{s}$,
where we have included a 2\% correction in $\Oex$ owing to the
metallic electrodes beneath the ions (Methods Summary).

With currently achieved size scales in ion traps, the Coulomb
interaction is relatively weak, so low ion heating rates and
stable trapping potentials are essential. Heating rates can be
suppressed by operating at cryogenic
temperatures~\cite{deslauriers06,labaziewicz08_2}, so that the direct
Coulomb coupling rate can exceed the heating rate. In this work, the
trap electrodes and surrounding vacuum enclosure are cooled to 4.2~K
with a liquid helium bath cryostat. With similar versions of this trap
at $\on/2\pi=2.3$~MHz, heating rates expressed as $d\langle n
\rangle/dt$ (where $n$ denotes the quantum number of motional Fock
state $|n\rangle$ and $\langle n \rangle$ is its expectation value)
were observed to be as low as 70~quanta per second, consistent with the results
of ref.~\citenum{labaziewicz08_2}. However, for the experiments
described here ($\on/2\pi\approx 4.0-5.6$~MHz) the heating rate was
$\sim500-2,000$~quanta per second and varied between the two wells.  We
observed $d\langle n \rangle/dt \propto 1/\omega_0^2$ in this trap, in
agreement with previous
reports~\cite{deslauriers06,labaziewicz08_2,epstein07}, so large
values of $\omega_0$ are beneficial. The use of $^9\mathrm{Be}^+$, the
lightest of the commonly trapped atomic ions, is an advantage here,
because for given d.c. trapping potentials the heating rate should
remain unchanged while $\Oex \propto m^{-1/2}$.  Cryogenic operation
decreases the background gas pressure to negligible levels, such that
ion loss rates due to collisions with background gas are smaller than
one per day.

\begin{figure}
\centering \includegraphics{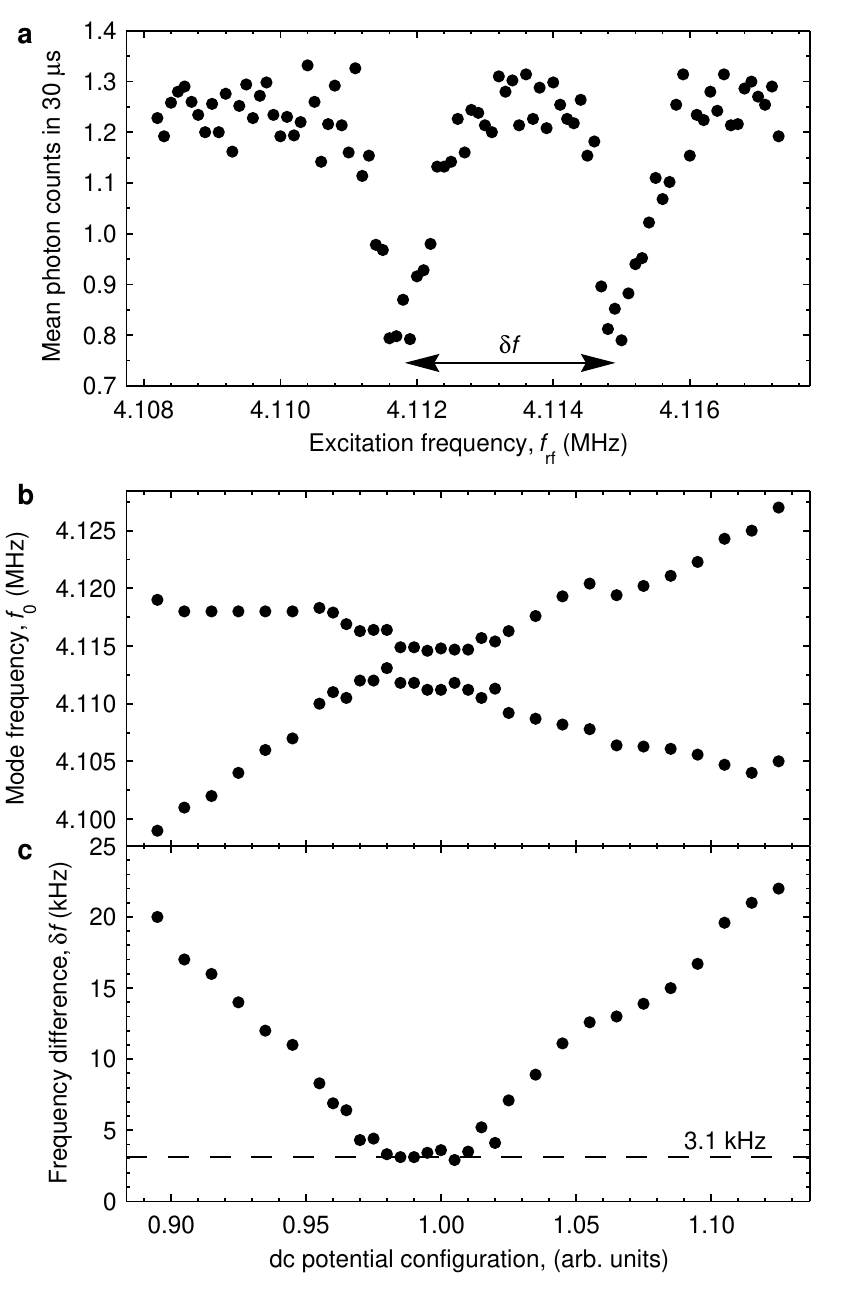}
  \sffamily \caption{Motional spectroscopy of two coupled ions near
  the avoided crossing. \textbf{a}, Decreases in collected
  fluorescence occur at values of excitation frequency,
  $f_\mathrm{rf}$, corresponding to the ion mode frequencies.  With
  $\tau_p = 960$~$\mu$s, the splitting on resonance is
  resolved. \textbf{b}, \textbf{c}, Mode frequencies (\textbf{b}) and
  mode frequency splitting, $\delta f$ (\textbf{c}), for the axial
  normal modes of two ions separated by 40~$\mu$m. Error bars are
  smaller than the size of the points. The data were acquired over a
  1-h period, and slow variations in ambient potentials gave rise
  to the fluctuations.}
\label{fig:avoided_crossing}
\end{figure}

A signature of coupling between the ions is the splitting between the
two axial normal mode frequencies. As the trap potential is tuned into
the resonance condition, this splitting, $\delta f$, reaches a
theoretical minimum $\delta f=\Oex/\pi=3.1$~kHz. A plot of the mode
frequencies will therefore show an avoided crossing. We measure the
mode frequencies by applying a nearly resonant oscillating potential
pulse to one of the trap electrodes. We then illuminate both ions with
laser radiation resonant with the $^2S_{1/2}-{^2P}_{3/2}$ cycling
transition at 313~nm. A decrease in the resulting fluorescence
indicates that a mode of the ions' motion has been resonantly
excited. For pulse lengths $\tau_p \gg 1/\delta f$, we resolve the two
modes (Fig.~\ref{fig:avoided_crossing}a).
We sweep the trapping
wells through resonance by varying the static potentials that are
applied to the trap electrodes. A plot of the resulting mode
frequencies, determined as above, is given in
Fig.~\ref{fig:avoided_crossing}b, c, showing a minimum of $\delta
f=3.0(5)$~kHz, in agreement with theory.

To demonstrate coupling at the level of a few motional quanta, we
observe the exchange of energy between the two ions as follows. The
ion motional frequencies are initially detuned by $100$~kHz, which is
much greater than $\Omega_\mathrm{ex}/(2\pi)$, effectively decoupling
the ions' motions. The ions are then simultaneously illuminated with a
laser detuned by $-10$~MHz from the $^2S_{1/2} - {^2P}_{3/2}$ cycling
transition, cooling them into a thermal state at the Doppler limit
with mean occupation~$\langle n \rangle = 2.3(1)$. Subsequently,
ion~$a$ is cooled to $\langle n_a \rangle = 0.35(2)$ by several cycles
of stimulated Raman cooling with the
$|\downarrow\rangle\equiv|F=2,m_F=-2\rangle$ and
$|\uparrow\rangle\equiv|1,-1\rangle$ hyperfine
states~\cite{monroe95}. The Raman beams are counter-propagating and
oriented at 45\textdegree\ relative to the $x$ axis. At this point the
potentials are brought into resonance ($\on/2\pi=4.04$~MHz) within
an interval (9~$\mu$s) short in comparison with $\tex$ but long in
comparison with the axial oscillation period. They remain on resonance
for a time $\tau$, allowing energy to transfer between the
ions. After a time $\tau$, the potentials are adiabatically returned
to their off-resonance values and we determine the mean quantum number,
$\langle n_a \rangle$, in ion~$a$ by observing asymmetry between the
red and blue motional sidebands of the $\ketdown$-to-$\ketup$
hyperfine Raman transition~\cite{monroe95}.

\begin{figure}
\centering \includegraphics{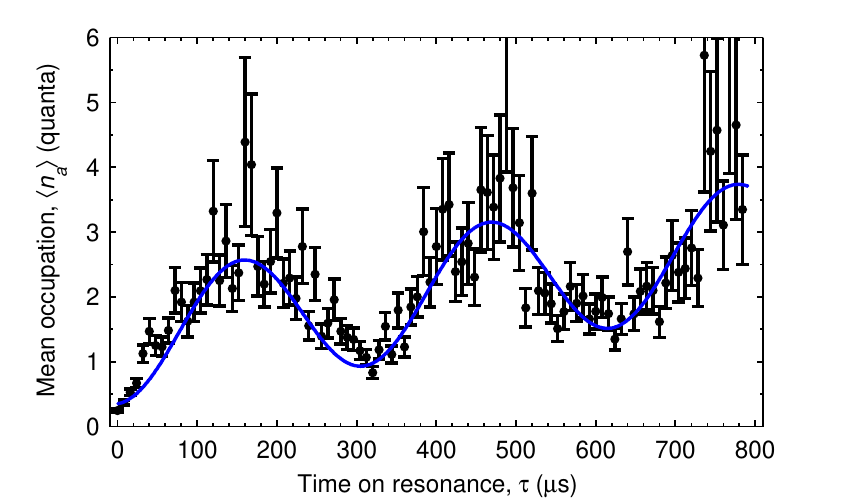}
  \sffamily \caption{Energy swapping between two ions in separate
  trapping potentials at the level of a few quanta. The mean
  occupation, $\langle n_a \rangle$, of ion~$a$ is plotted with error
  bars (s.e.m.) for various durations, $\tau$, that the ion motional
  frequencies remain on resonance. The blue curve represents a fit to
  theory with four free parameters: the two initial mean quantum
  numbers, the exchange time and the heating rate. Energy exchanges
  between the ions at 155(1)-$\mu$s intervals. The linearly increasing
  trend in $\langle n_a \rangle$ is due to ion heating at a rate of
  1,885(10) quanta per second. Uncertainties represent standard errors
  of the fit parameters.}
\label{fig:thermal_exchange}
\end{figure}

As seen in Fig.~\ref{fig:thermal_exchange}, energy exchanges between
the ions during an interval $\tex=155(1)$~$\mu$s.
The 5\% disagreement between the measured and predicted (162~$\mu$s)
values for $\tex$ is probably due to uncertainty in the ion
separation, $s_0$ (even a 1-$\mu$m uncertainty would account for the
disagreement). The first maximum of $\langle n_a \rangle$ corresponds
to the cooling limit of ion~$b$ ($\langle
n_b(\tau=0) \rangle=2.3(1)$~quanta).  The underlying linear growth in
$\langle n_a \rangle$ corresponds to heating of the ions at a rate of
$\dot{\bar{n}} = 1,885(10)$ quanta per second (Methods Summary).

\begin{figure}
\centering \includegraphics{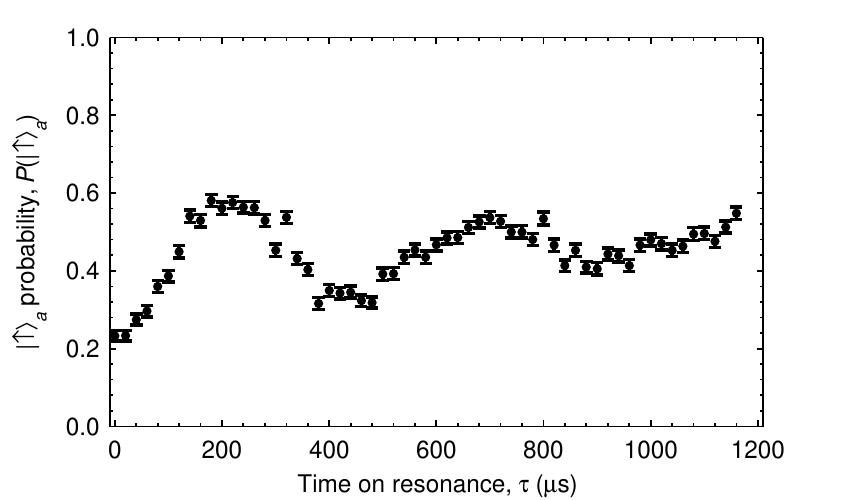}
  \sffamily
  \caption{Motional exchange between two ions in separate
    trapping potentials at approximately the single-quantum level. The
    probability, $P(|\uparrow\rangle_a)$, of measuring ion~$a$ in spin
    state $|\uparrow\rangle_a$ at the end of the experimental sequence
    is plotted with error bars (s.e.m.) against the time,
    $\tau$, for which the ions interact. $P(|\uparrow\rangle_a)$ oscillates
    with period $2\tex = 437(4)$~$\mu$s as a quantum exchanges between the
    ions.}
\label{fig:quantum_exchange}
\end{figure}

As a final experiment, we demonstrate energy exchange at approximately
the single-quantum level. Ideally, the experiment takes the following
form. The ions are tuned to the resonance condition throughout and are
initially Doppler cooled. Ion~$a$ is Raman-cooled, sympathetically
cooling ion~$b$ and thereby preparing the state
$|0\rangle_a|\downarrow\rangle_a|0\rangle_b$. To create a single
motional quantum, we drive ion~$a$ with a blue-sideband Raman $\pi$
pulse (of duration $10$~$\mu$s, which is much less than $\tex$),
creating the state $|1\rangle_a|\uparrow\rangle_a|0\rangle_b$. The
system oscillates between $|1\rangle_a|\uparrow\rangle_a|0\rangle_b$
and $|0\rangle_a|\uparrow\rangle_a|1\rangle_b$ with period
$2\tex$. After a time $\tau$, we drive ion~$a$ with another
blue-sideband $\pi$ pulse, conditionally flipping the spin from
$|1\rangle_a|\uparrow\rangle_a$ to $|0\rangle_a|\downarrow\rangle_a$,
dependent on the presence of a motional quantum in ion~$a$. The final
internal state probability will be given by
$P(|\uparrow\rangle_a)(\tau)=\sin^2(\Oex \tau)$. In practice, contrast
in the oscillations (Fig.~\ref{fig:quantum_exchange}) is
significantly reduced by incomplete cooling, motional decoherence and
decoherence due to imperfect Raman sideband pulses.
We estimate that ion~$a$ is cooled initially to $\langle n_a \rangle
=0.3(1)$. Although we were unable to measure the initial temperature
of ion~$b$ directly, comparison of the contrast and temporal behavior
of our exchange data (Fig.~\ref{fig:quantum_exchange}) with
simulations indicates that $\langle n_b \rangle \lesssim
0.6$. Motional decoherence results from heating and from trap
frequency instability over the time required to acquire the
data. Raman sideband pulses suffer from variations in laser intensity
and fluctuations in the sideband coupling caused by thermal spread in
the $y$ and $z$ motional states (Debye-Waller
factors~\cite{wineland98}). For $\omega_0/2\pi=5.56$~MHz, we observe
oscillations with period $2\tex=437(4)$~$\mu$s. The 2\% disagreement
with the prediction (447~$\mu$s) of
equations \ref{eqn:interaction_strength} and \ref{eqn:sol} is probably
due to uncertainty in the ion separation and the difficulty of
maintaining the exact resonance condition.

Significant improvements in coupling fidelity seem to be in reach
with current technology. It should be possible to improve Raman laser
intensity stability, and Debye-Waller factors from the $y$ and $z$
motion can be eliminated by proper choice of beam
directions~\cite{monroe95}. We believe that our motional frequency
instability can be alleviated by ensuring that the trap surface is
free of charged contaminants.  Faster exchange can be achieved by
scaling down trap dimensions, but this puts a premium on reduced
motional heating. Although the work presented here uses the axial mode
to couple the ions, it may prove advantageous to use the radial modes
in future experiments because of the lower heating rates associated
with their higher frequencies. As an example, radial mode frequencies of
$\sim30$~MHz are routinely achieved in the apparatus, compared with
axial frequencies $\lesssim10$~MHz.

These results could lead to several possible applications in quantum
state engineering and spectroscopy. For example, from the motional
state $|1\rangle_a|0\rangle_b$ the state of the ions at time
$t=\tex/2$ is the Bell state $(|1\rangle_a|0\rangle_b +
i|0\rangle_a|1\rangle_b)/\sqrt{2}$. Transferring this state onto the
ions' internal states with sideband pulses would create an entangled
spin state, even between ions of dissimilar species. This could be
used as an entangled-pair factory in the scheme of
refs~\citenum{wineland98},\citenum{kielpinski02}, with the
advantage over previous schemes~\cite{jost09} that the ions are
already in separate wells, ready for distribution to separate
locations. The coupling could be used to read out the state of one ion
species with another, an ability useful for error correction protocols
and for quantum logic spectroscopy~\cite{schmidt05}. When the harmonic
wells are not in resonance, the spin state of one ion can be read out
without destroying the state of the other, so schemes for weak or
quantum non-demolition measurements that use either Kerr-type
nonlinearities or quantum logic may become
feasible~\cite{haroche06}. Hybrid quantum systems, incorporating
similar interactions to couple a trapped ion to other quantum devices,
could serve as a means of transferring quantum information between
different qubit implementations in a future quantum network. For
example, a trapped ion could act as a quantum transformer between a
superconducting qubit~\cite{tian05} and a photonic
qubit~\cite{cirac97,moehring07}. The sympathetic cooling through
exchange might also be used to cool neutral
molecules~\cite{idziaszek10}.

\vspace{3 mm}
\noindent \textbf{METHODS SUMMARY}

\noindent The shielding factor, $\beta$, represents the ratio of
the exchange rates $\Oex$ with and without the presence of the
trapping electrodes. To a good approximation, we ensure that at the
motional frequencies all trap electrodes are held at
ground. Therefore, assuming that gaps between the electrodes are
negligible, the shielding factor can be calculated with the method of
images. The result is
\begin{equation}
\beta=1 - \frac{1}{2}\left(\frac{3 (s_0/d_0)^5}{\left(4 +(s_0/d_0)^2\right)^{5/2}}-\frac{(s_0/d_0)^3}{\left(4 + (s_0/d_0)^2\right)^{3/2}}\right)
\end{equation}
which reaches a maximum of $\beta=1.018$ at $s_0=d_0$.

The evolution of $\langle n_a \rangle$ under the
influence of equation~\ref{eqn:interaction}, including heating effects and
assuming that both ions begin in a thermal state with mean quantum
numbers $n_{a0}$ and $n_{b0}$, can be predicted with a Langevin
equation~\cite{heinzen90}. On resonance, the evolution is
\begin{equation}
\langle n_a \rangle = n_{a0} \cos^2 (\Oex t) + n_{b0} \sin^2 (\Oex t) + \dot{\bar{n}} t
\end{equation}
where $\dot{\bar{n}}$ represents the mean of $d\langle n_a \rangle/dt$
and $d\langle n_b \rangle/dt$ for uncorrelated noise sources.

\bibliography{kenton}

\begin{thebibliography}{10}
\expandafter\ifx\csname url\endcsname\relax
  \def\url#1{\texttt{#1}}\fi
\expandafter\ifx\csname urlprefix\endcsname\relax\def\urlprefix{URL }\fi
\providecommand{\bibinfo}[2]{#2}
\providecommand{\eprint}[2][]{\url{#2}}

\bibitem{haroche06}
\bibinfo{author}{Haroche, S.} \& \bibinfo{author}{Raimond, J.-M.}
\newblock \emph{\bibinfo{title}{Exploring the Quantum: Atoms, Cavities, and
  Photons}} (\bibinfo{publisher}{Oxford Univ. Press}, \bibinfo{year}{2006}).

\bibitem{miller05}
\bibinfo{author}{Miller, R.} \emph{et~al.}
\newblock \bibinfo{title}{Trapped atoms in cavity {QED}: coupling quantized
  light and matter}.
\newblock \emph{\bibinfo{journal}{J. Phys. B}} \textbf{\bibinfo{volume}{38}},
  \bibinfo{pages}{S551} (\bibinfo{year}{2005}).

\bibitem{houck07}
\bibinfo{author}{Houck, A.~A.} \emph{et~al.}
\newblock \bibinfo{title}{Generating single microwave photons in a circuit}.
\newblock \emph{\bibinfo{journal}{Nature}} \textbf{\bibinfo{volume}{449}},
  \bibinfo{pages}{328--331} (\bibinfo{year}{2007}).

\bibitem{leibfried03}
\bibinfo{author}{Leibfried, D.}, \bibinfo{author}{Blatt, R.},
  \bibinfo{author}{Monroe, C.} \& \bibinfo{author}{Wineland, D.~J.}
\newblock \bibinfo{title}{Quantum dynamics of single trapped ions}.
\newblock \emph{\bibinfo{journal}{Rev. Mod. Phys.}}
  \textbf{\bibinfo{volume}{75}}, \bibinfo{pages}{281--324}
  (\bibinfo{year}{2003}).

\bibitem{oconnell10}
\bibinfo{author}{O'Connell, A.~D.} \emph{et~al.}
\newblock \bibinfo{title}{Quantum ground state and single-phonon control of a
  mechanical resonator}.
\newblock \emph{\bibinfo{journal}{Nature}} \textbf{\bibinfo{volume}{464}},
  \bibinfo{pages}{697--703} (\bibinfo{year}{2010}).

\bibitem{monroe01}
\bibinfo{author}{Monroe, C.} \emph{et~al.}
\newblock \bibinfo{title}{Scalable entanglement of trapped ions}.
\newblock In \bibinfo{editor}{Arimondo, E.}, \bibinfo{editor}{De~Natale, P.} \&
  \bibinfo{editor}{Inguscio, M.} (eds.) \emph{\bibinfo{booktitle}{Atomic
  Physics 17, Proceedings of the 17th International Conference}},
  \bibinfo{pages}{173--186} (\bibinfo{year}{2001}).

\bibitem{jost09}
\bibinfo{author}{Jost, J.~D.} \emph{et~al.}
\newblock \bibinfo{title}{Entangled mechanical oscillators}.
\newblock \emph{\bibinfo{journal}{Nature}} \textbf{\bibinfo{volume}{459}},
  \bibinfo{pages}{683--685} (\bibinfo{year}{2009}).

\bibitem{harlander11}
\bibinfo{author}{Harlander, M.}, \bibinfo{author}{Lechner, R.},
  \bibinfo{author}{Brownnutt, M.}, \bibinfo{author}{Blatt, R.} \&
  \bibinfo{author}{H\"{a}nsel, W.}
\newblock \bibinfo{title}{Trapped-ion antennae for the transmission of quantum
  information}.
\newblock \emph{\bibinfo{journal}{Nature}} \textbf{\bibinfo{volume}{471}},
  \bibinfo{pages}{200--203} (\bibinfo{year}{2011}).

\bibitem{heinzen90}
\bibinfo{author}{Heinzen, D.~J.} \& \bibinfo{author}{Wineland, D.~J.}
\newblock \bibinfo{title}{Quantum-limited cooling and detection of
  radio-frequency oscillations by laser-cooled ions}.
\newblock \emph{\bibinfo{journal}{Phys. Rev. A}} \textbf{\bibinfo{volume}{42}},
  \bibinfo{pages}{2977--2994} (\bibinfo{year}{1990}).

\bibitem{wineland98}
\bibinfo{author}{Wineland, D.~J.} \emph{et~al.}
\newblock \bibinfo{title}{Experimental issues in coherent quantum-state
  manipulation of trapped atomic ions}.
\newblock \emph{\bibinfo{journal}{J. Res. Nat. Inst. Stand. Tech.}}
  \textbf{\bibinfo{volume}{103}}, \bibinfo{pages}{259--328}
  (\bibinfo{year}{1998}).

\bibitem{tian04}
\bibinfo{author}{Tian, L.} \& \bibinfo{author}{Zoller, P.}
\newblock \bibinfo{title}{Coupled ion-nanomechanical systems}.
\newblock \emph{\bibinfo{journal}{Phys. Rev. Lett.}}
  \textbf{\bibinfo{volume}{93}}, \bibinfo{pages}{266403}
  (\bibinfo{year}{2004}).

\bibitem{hensinger05}
\bibinfo{author}{Hensinger, W.~K.} \emph{et~al.}
\newblock \bibinfo{title}{Ion trap transducers for quantum electromechanical
  oscillators}.
\newblock \emph{\bibinfo{journal}{Phys. Rev. A}} \textbf{\bibinfo{volume}{72}},
  \bibinfo{pages}{041405(R)} (\bibinfo{year}{2005}).

\bibitem{tian05}
\bibinfo{author}{Tian, L.}, \bibinfo{author}{Blatt, R.} \&
  \bibinfo{author}{Zoller, P.}
\newblock \bibinfo{title}{Scalable ion trap quantum computing without moving
  ions}.
\newblock \emph{\bibinfo{journal}{Eur. Phys. J. D}}
  \textbf{\bibinfo{volume}{32}}, \bibinfo{pages}{201--208}
  (\bibinfo{year}{2005}).

\bibitem{schmied08}
\bibinfo{author}{Schmied, R.}, \bibinfo{author}{Roscilde, T.},
  \bibinfo{author}{Murg, V.}, \bibinfo{author}{Porras, D.} \&
  \bibinfo{author}{Cirac, J.~I.}
\newblock \bibinfo{title}{Quantum phases of trapped ions in an optical
  lattice}.
\newblock \emph{\bibinfo{journal}{New J. Phys.}} \textbf{\bibinfo{volume}{10}},
  \bibinfo{pages}{045017} (\bibinfo{year}{2008}).

\bibitem{chiaverini08}
\bibinfo{author}{Chiaverini, J.} \& \bibinfo{author}{{Lybarger, Jr.}, W.~E.}
\newblock \bibinfo{title}{Laserless trapped-ion quantum simulations without
  spontaneous scattering using microtrap arrays}.
\newblock \emph{\bibinfo{journal}{Phys. Rev. A}} \textbf{\bibinfo{volume}{77}},
  \bibinfo{pages}{022324} (\bibinfo{year}{2008}).

\bibitem{schmied09}
\bibinfo{author}{Schmied, R.}, \bibinfo{author}{Wesenberg, J.~H.} \&
  \bibinfo{author}{Leibfried, D.}
\newblock \bibinfo{title}{Optimal surface-electrode trap lattices for quantum
  simulation with trapped ions}.
\newblock \emph{\bibinfo{journal}{Phys. Rev. Lett.}}
  \textbf{\bibinfo{volume}{102}}, \bibinfo{pages}{233002}
  (\bibinfo{year}{2009}).

\bibitem{cirac00}
\bibinfo{author}{Cirac, J.~I.} \& \bibinfo{author}{Zoller, P.}
\newblock \bibinfo{title}{A scalable quantum computer with ions in an array of
  microtraps}.
\newblock \emph{\bibinfo{journal}{Nature}} \textbf{\bibinfo{volume}{404}},
  \bibinfo{pages}{579--581} (\bibinfo{year}{2000}).

\bibitem{kielpinski02}
\bibinfo{author}{Kielpinski, D.}, \bibinfo{author}{Monroe, C.} \&
  \bibinfo{author}{Wineland, D.~J.}
\newblock \bibinfo{title}{Architecture for a large-scale ion-trap quantum
  computer}.
\newblock \emph{\bibinfo{journal}{Nature}} \textbf{\bibinfo{volume}{417}},
  \bibinfo{pages}{709--711} (\bibinfo{year}{2002}).

\bibitem{schmidt05}
\bibinfo{author}{Schmidt, P.~O.} \emph{et~al.}
\newblock \bibinfo{title}{Spectroscopy using quantum logic}.
\newblock \emph{\bibinfo{journal}{Science}} \textbf{\bibinfo{volume}{309}},
  \bibinfo{pages}{749--752} (\bibinfo{year}{2005}).

\bibitem{rosenband08}
\bibinfo{author}{Rosenband, T.} \emph{et~al.}
\newblock \bibinfo{title}{Frequency ratio of {Al$^+$} and {Hg$^+$} single-ion
  optical clocks; metrology at the 17th decimal place}.
\newblock \emph{\bibinfo{journal}{Science}} \textbf{\bibinfo{volume}{319}},
  \bibinfo{pages}{1808--1812} (\bibinfo{year}{2008}).

\bibitem{daniilidis09}
\bibinfo{author}{Daniilidis, N.}, \bibinfo{author}{Lee, T.},
  \bibinfo{author}{Clark, R.}, \bibinfo{author}{Narayanan, S.} \&
  \bibinfo{author}{H\"{a}ffner, H.}
\newblock \bibinfo{title}{Wiring up trapped ions to study aspects of quantum
  information}.
\newblock \emph{\bibinfo{journal}{J. Phys. B}} \textbf{\bibinfo{volume}{42}},
  \bibinfo{pages}{154012} (\bibinfo{year}{2009}).

\bibitem{tan02}
\bibinfo{author}{Tan, J.~N.}
\newblock \bibinfo{title}{Interacting ion oscillators in contiguous confinement
  wells}.
\newblock \emph{\bibinfo{journal}{Bull. Am. Phys. Soc.}}
  \textbf{\bibinfo{volume}{47}}, \bibinfo{pages}{103} (\bibinfo{year}{2002}).

\bibitem{ciaramicoli03}
\bibinfo{author}{Ciaramicoli, G.}, \bibinfo{author}{Marzoli, I.} \&
  \bibinfo{author}{Tombesi, P.}
\newblock \bibinfo{title}{Scalable quantum processor with trapped electrons}.
\newblock \emph{\bibinfo{journal}{Phys. Rev. Lett.}}
  \textbf{\bibinfo{volume}{91}}, \bibinfo{pages}{017901}
  (\bibinfo{year}{2003}).

\bibitem{seidelin06}
\bibinfo{author}{Seidelin, S.} \emph{et~al.}
\newblock \bibinfo{title}{Microfabricated surface-electrode ion trap for
  scalable quantum information processing}.
\newblock \emph{\bibinfo{journal}{Phys. Rev. Lett.}}
  \textbf{\bibinfo{volume}{96}}, \bibinfo{pages}{253003}
  (\bibinfo{year}{2006}).

\bibitem{deslauriers06}
\bibinfo{author}{Deslauriers, L.} \emph{et~al.}
\newblock \bibinfo{title}{Scaling and suppression of anomalous heating in ion
  traps}.
\newblock \emph{\bibinfo{journal}{Phys. Rev. Lett.}}
  \textbf{\bibinfo{volume}{97}}, \bibinfo{pages}{103007}
  (\bibinfo{year}{2006}).

\bibitem{labaziewicz08_2}
\bibinfo{author}{Labaziewicz, J.} \emph{et~al.}
\newblock \bibinfo{title}{Temperature dependence of electric field noise above
  gold surfaces}.
\newblock \emph{\bibinfo{journal}{Phys. Rev. Lett.}}
  \textbf{\bibinfo{volume}{101}}, \bibinfo{pages}{180602}
  (\bibinfo{year}{2008}).

\bibitem{epstein07}
\bibinfo{author}{Epstein, R.~J.} \emph{et~al.}
\newblock \bibinfo{title}{Simplified motional heating rate measurements of
  trapped ions}.
\newblock \emph{\bibinfo{journal}{Phys. Rev. A}} \textbf{\bibinfo{volume}{76}},
  \bibinfo{pages}{033411} (\bibinfo{year}{2007}).

\bibitem{monroe95}
\bibinfo{author}{Monroe, C.} \emph{et~al.}
\newblock \bibinfo{title}{Resolved-sideband {R}aman cooling of a bound atom to
  the 3{D} zero-point energy}.
\newblock \emph{\bibinfo{journal}{Phys. Rev. Lett.}}
  \textbf{\bibinfo{volume}{75}}, \bibinfo{pages}{4011--4014}
  (\bibinfo{year}{1995}).

\bibitem{cirac97}
\bibinfo{author}{Cirac, J.~I.}, \bibinfo{author}{Zoller, P.},
  \bibinfo{author}{Kimble, H.~J.} \& \bibinfo{author}{Mabuchi, H.}
\newblock \bibinfo{title}{Quantum state transfer and entanglement distribution
  among distant nodes in a quantum network}.
\newblock \emph{\bibinfo{journal}{Phys. Rev. Lett.}}
  \textbf{\bibinfo{volume}{78}}, \bibinfo{pages}{3221--3224}
  (\bibinfo{year}{1997}).

\bibitem{moehring07}
\bibinfo{author}{Moehring, D.~L.} \emph{et~al.}
\newblock \bibinfo{title}{{Entanglement of single-atom quantum bits at a
  distance}}.
\newblock \emph{\bibinfo{journal}{{Nature}}} \textbf{\bibinfo{volume}{{449}}},
  \bibinfo{pages}{{68--71}} (\bibinfo{year}{{2007}}).

\bibitem{idziaszek10}
\bibinfo{author}{Idziaszek, Z.}, \bibinfo{author}{Calarco, T.} \&
  \bibinfo{author}{Zoller, P.}
\newblock \bibinfo{title}{Ion-assisted ground-state cooling of a trapped polar
  molecule}.
\newblock \emph{\bibinfo{journal}{http://arxiv.org/abs/1008.1858}}
  (\bibinfo{year}{2010}).

\end{thebibliography}

\noindent\textbf{Acknowledgements}
This work was supported by IARPA, DARPA, ONR and the NIST Quantum
  Information Program. We thank M. Biercuk, A. VanDevender, J. Amini,
  and R. B. Blakestad for their help in assembling parts of the
  experiment, and we thank U. Warring and R. Simmonds for
  comments. This paper, a submission of NIST, is not subject to US
  copyright.

\noindent\textbf{Author Contributions}
K.R.B. and C.O. participated in the
  design of the experiment and built the experimental
  apparatus. K.R.B. collected data, analysed results and wrote the
  manuscript. Y.C. fabricated the ion trap chip and collected
  data. A.C.W. maintained laser systems and collected
  data. D.L. participated in the design of the experiment,
  collected data and maintained laser systems. D.J.W. participated in
  the design and analysis of the experiment. All authors discussed
  the results and the text of the manuscript.

\noindent\textbf{Author Information}
Reprints and permissions information is
  available at www.nature.com/reprints. The authors declare no
  competing financial interests. Readers are welcome to comment on the
  online version of this article at
  www.nature.com/nature. Correspondence and requests for materials
  should be addressed to K.R.B. (\mbox{kenton.brown@nist.gov}).

\end{document}